# Fourier's quantum information processing

Mario Mastriani
*School of Computing & Information Sciences, Florida International University,*
*11200 S.W. 8th Street, Miami, FL 33199, USA*
mario.mastriani@fiu.edu

We demonstrate that quantum information processing (QIP) completely rests on quantum Fourier transform (QFT), and 190 years after his death, the work of Jean-Baptiste Joseph Fourier is more present than ever in Physics, constituting the heart of QIP, and showing the spectral nature of quantum entanglement, quantum teleportation, and quantum secret sharing.

## I. INTRODUCTION

The fact that one hundred and ninety years after his death, the work of Jean-Baptiste Joseph Fourier (21 March 1768 - 16 May 1830) continues to be the center of the following scientific disciplines:
- All heat Physics,
- Optics,
- circuit analysis in Electricity,
- pulse and wideband analysis in Electronics,
- Computing,
- a type of digital transmission and a method of encoding digital data on multiple carrier frequencies called orthogonal frequency-division multiplexing (OFDM) in telecommunications, which has developed into a popular scheme for wideband digital communication used in applications such as digital television and audio broadcasting, digital subscriber line (DSL) internet access, wireless networks, power line networks, and 4G/5G mobile communications,
- Automatic Control,
- sound propagation in solid materials,
- convolution in synthetic aperture radar via Schönhage–Strassen algorithm,
- filtering and compression in Digital Signal, Image and Video Processing,
- protein docking, Fourier Transform Ion Cyclotron Resonance, correlation in genomic, proteomic, and metabolomics in Molecular Biology, as well as, high speed multiple sequence alignment, and Fast Fourier Transform-Based Correlation of Deoxyribonucleic acid (DNA) Sequences Using Complex Plane Encoding in Bioinformatics,
- spot intensities measurement and image generation by Fourier transformation of the intensities in X-Ray Crystallography,
- and so on,

does not represent anything new to anyone. What is surprising is its timeless projection on the most disruptive of all technologies currently under scrutiny by the scientific community, i.e., Quantum Information Processing (QIP) [1-3]. In fact, the inexcusable dependency that QIP has on Quantum Fourier Transform (QFT) [1, 4-6] is clearly highlighted in this work. This dependence will be evident both for individual gates of one to four qubits, as well as in those cases where more complex configurations must be implemented, such as the case of the most notable and at the same time strange effect of the QIP, i.e., the entanglement [7-9] between two, three, four or more qubits, as well as the projection of this effect on its most outstanding application, known as quantum teleportation [10-13].

On the other hand, it is well known that QFT is key in the phase detection of innumerable quantum algorithms like that of Shor [14], however, it is not so well known that QFT is behind the nature and consequence of the flip errors and noises present in every QIP experiment [1]. Quantum gates, particularly those optically implemented [15], inherently generate a series of noises, specifically, three types: bit flip, phase flip (or phase damping), and bit-phase flip, where undoubtedly, the influence of these noises on total process performance is in direct proportion to the number of gates used by the quantum algorithm on every qubit.

To summarize, this work allows us to establish that all relative aspects of QIP are covered by QFT, representing its most important tool for its finer analysis and future expansion.

We hope this work is of interest to theorists, experimentalists, and others who might want to study the underlying relationships behind all quantum algorithms, their respective implementations on a physical machine (based on superconductor technology [16, 17]) or an optical circuit [15], and an in-depth analysis of the respective outcomes. In Section II, all equivalences between every quantum gate, from one to four qubits, and their respective representation via QFT are developed. Section III is especially dedicated to entanglement [7-9] for Greenberger–Horne–Zeilinger (GHZ) [1] states of two ($GHZ_2$), three ($GHZ_3$), and four ($GHZ_4$) entangled particles; and quantum teleportation [10-13]. Finally, Section IV has our conclusions.

## II. FOURIER'S QUANTUM INFORMATION PROCESSING

In this section, we will establish the equivalences between Quantum Fourier Transform (QFT) and all the gates involved in Quantum Information Processing (QIP). In fact, we will analyze these equivalences from least to greatest, i.e., for gates from one to four qubits, for which, we will start with a brief description of the QFT and its inverse.

The QFT represents an important family of quantum operations over the ring $\mathbb{Z}_2^n$. The $n$-qubit QFT coherently transforms an input state $|x\rangle$ in the computational basis as follows [18]:

$$|x\rangle \mapsto \frac{1}{\sqrt{2^n}} \sum_{v=0}^{2^n-1} \omega_{2^n}^{u.v} |y\rangle, \qquad u = 0, 1, 2, \ldots, 2^n - 1, \tag{1}$$

where $|x\rangle$ represents a qubit-string $|x_1 \ldots x_n\rangle$, and $\omega_{2^n} = e^{i2\pi/2^n}$ is the $2^n$ *root of unity*, such that $\omega_{2^n}^{2^n} = 1$, while the inverse QFT is:

$$|y\rangle \mapsto \frac{1}{\sqrt{2^n}} \sum_{u=0}^{2^n-1} \omega_{2^n}^{-v.u} |x\rangle, \qquad v = 0, 1, 2, \ldots, 2^n - 1. \tag{2}$$

Then, if $F_{2^n}$ represents the matrix of QFT [1] and $F_{2^n}^{-1}$ its reverse, the simple number *1* (unity), i.e., QFT for *0*-qubits, $F_{2^0} = F_{2^0}^{-1} = 1$, will be present, *per se*, inside all the gates used in QIP [1].

### A. Equivalences for *1*-qubit gates

It is known from literature [19, 20], that Hadamard matrix $H$ is equivalent to the 2-qubit QFT and its inverse, that is,

$$F_{2^1} = H = \frac{1}{\sqrt{2}} \begin{bmatrix} 1 & 1 \\ 1 & -1 \end{bmatrix} = H^{-1} = F_{2^1}^{-1}. \tag{3}$$

What is not so present in the literature is that both the Pauli's and phase matrices [1] are derived from the Hadamard matrix $H$ through simple arithmetic and flipping operations. In fact, if we flip over the matrix $H$ with respect to an imaginary horizontal axis that crosses it in the middle, see Fig. 1, we will obtain the splitting operator in its matrix form [21]. Then, the splitting operator together with the recombining operator, which arises from doing the same thing that we have done in Fig. 1 but with respect to an imaginary vertical axis that crosses halfway through the matrix $H$, constitute the pair of operators used in the interference experiments [21].

   *1. Pauli's matrices.* Next, we will establish the existing relations between the Pauli's matrices [1] (*I*: identity, *X*: inverter, *Y*, and *Z*) and the Hadamard matrix *H* from simple arithmetic and flipping operations, thus establishing the underlying relation between the Pauli's matrices and the QFT.

$$H = \frac{1}{\sqrt{2}}\begin{bmatrix} 1 & 1 \\ 1 & -1 \end{bmatrix} \xrightarrow{\text{flipping the matrix}} H_{\text{flipped}} = \frac{1}{\sqrt{2}}\begin{bmatrix} 1 & -1 \\ 1 & 1 \end{bmatrix}$$

FIG. 1. Flipping the Hadamard matrix $H$ with respect to an imaginary horizontal axis that crosses it in the middle, where $H_{\text{flipped}} = X H$, and $X$ is the inverter Pauli matrix, or flipping matrix.

$$I = HH = \begin{bmatrix} 1 & 0 \\ 0 & 1 \end{bmatrix} = XX, \tag{4a}$$

$$X = I_{\text{flipped}} = X I = \begin{bmatrix} 0 & 1 \\ 1 & 0 \end{bmatrix}, \tag{4b}$$

$$Z = X\left(H_{\text{flipped}} H_{\text{flipped}}\right) = X\left((XH)(XH)\right) = HXH = \begin{bmatrix} 1 & 0 \\ 0 & -1 \end{bmatrix}, \tag{4c}$$

$$Y = i Z_{\text{flipped}} = i X Z = \begin{bmatrix} 0 & -i \\ i & 0 \end{bmatrix}, \tag{4d}$$

where $H_{\text{flipped}} = X H$ is the operation shown in Fig. 1 that flips the matrix $H$ with respect to an imaginary horizontal axis that crosses it in the middle, $Z_{\text{flipped}} = X Z$, $i = \sqrt{-1}$, and $H = H^{-1} = (X+Z)/\sqrt{2}$.

2. *Phase matrices* [1]. These matrices are derived from the $Z$ matrix of Eq.(4c), where,

$$S = \sqrt{Z} = \begin{bmatrix} 1 & 0 \\ 0 & i \end{bmatrix}, \tag{5a}$$

$$S^{-1} = conj(S) = \begin{bmatrix} 1 & 0 \\ 0 & -i \end{bmatrix}, \tag{5b}$$

$$T = \sqrt{S} = \begin{bmatrix} 1 & 0 \\ 0 & e^{i\frac{\pi}{4}} \end{bmatrix}, \tag{5c}$$

$$T^{-1} = conj(T) = \begin{bmatrix} 1 & 0 \\ 0 & e^{-i\frac{\pi}{4}} \end{bmatrix}, \tag{5d}$$

$$U = \sqrt{T} = \begin{bmatrix} 1 & 0 \\ 0 & e^{i\frac{\pi}{8}} \end{bmatrix}, \tag{5e}$$

$$U^{-1} = conj(U) = \begin{bmatrix} 1 & 0 \\ 0 & e^{-i\frac{\pi}{8}} \end{bmatrix}, \tag{5f}$$

where $conj(\bullet)$ means conjugate of "$\bullet$".

3. *Square-root-of-Not matrices*. These matrices are derived from the $X$ matrix of Eq.(4b), where,

$$V = \sqrt{X} = HTTH = \frac{(1+i)}{2}\begin{bmatrix} 1 & -i \\ -i & 1 \end{bmatrix}, \text{ and} \tag{6a}$$

$$V^{-1} = conj(V) = HT^{-1}T^{-1}H = \frac{(1+i)}{2}\begin{bmatrix} -i & 1 \\ 1 & -i \end{bmatrix}. \tag{6b}$$

Therefore, we have been able to verify here that all the 1-qubit matrices used in QIP derive from the Hadamard gate $H$, or QFT $F_{2^1}$, and will be used below for gates of more than 1-qubit.

### B. Equivalences for *2*-qubit gates

In the same way as in the previous subsection where we established a relationship between the QFT $F_{2^1}$ and the Hadamard matrix $H$ for 1-qubit, in this section, we will do the same between QFT of 2-qubits $F_{2^2}$, and the *CNOT* gate [1]. Then, being,

$$F_{2^2} = \frac{1}{2}\begin{bmatrix} 1 & 1 & 1 & 1 \\ 1 & i & -1 & -i \\ 1 & -1 & 1 & -1 \\ 1 & -i & -1 & i \end{bmatrix}, \quad (7a) \qquad \text{and} \qquad F_{2^2}^{-1} = \frac{1}{2}\begin{bmatrix} 1 & 1 & 1 & 1 \\ 1 & -i & -1 & i \\ 1 & -1 & 1 & -1 \\ 1 & i & -1 & -i \end{bmatrix}, \quad (7b)$$

it is evident that QFT and its inverse do not coincide for the 2-qubit case ($F_{2^2} \neq F_{2^2}^{-1}$) as in the previous cases. Therefore, we will proceed to represent both of them in Fig. 2 by means of *H*, Controlled-*S*, and *SWAP* gates for the case of QFT, and *H*, Controlled-$S^{-1}$, and *SWAP* gates for the case of QFT$^{-1}$. The difference is that Controlled-*S* is used in QFT, while Controlled-$S^{-1}$ is used in QFT$^{-1}$.

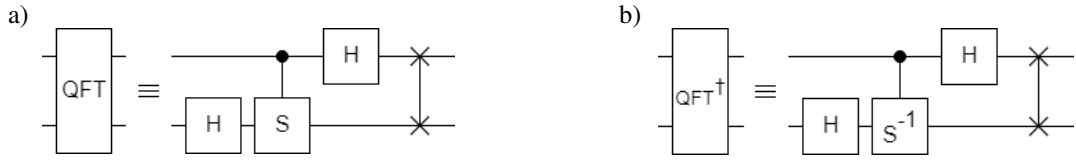

FIG. 2. a) QFT, $F_{2^2} \in \mathbb{C}^{2^2 \times 2^2}$ in terms of *H*, Controlled-*S*, and *SWAP* gates. b) Inverse of QFT, $F_{2^2}^{-1} \in \mathbb{C}^{2^2 \times 2^2}$ in terms of *H*, Controlled-$S^{-1}$, and *SWAP* gates.

Regardless of what was said above, it is necessary to establish with which of the two criteria present in Fig. 3 we will work from here on, mainly in relation to the Feynman's gate [1-3], also known as Controlled-*NOT*, *CNOT*, Controlled-*X*, or simply *CX*. Figure 3 shows two associated matrices to *CNOT* gate, depending on which qubit the control is, as well as, which qubit the target is [22]. Different books, papers, and quantum platforms [16, 17] order their qubits differently. In this paper, the option of Fig. 3(b) is chosen, precisely, because it is closer to that used in the main quantum platforms [16, 17].

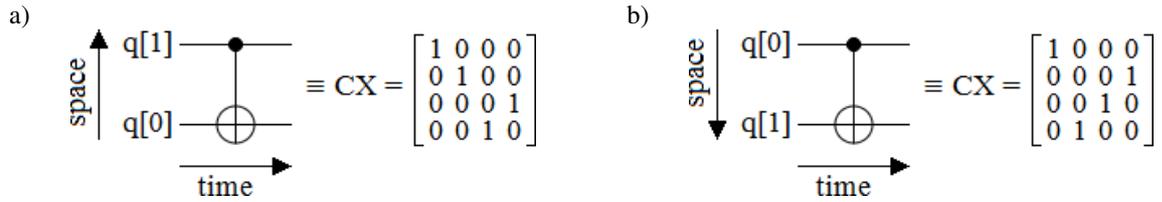

FIG. 3. Two options for the implementation of *CNOT* gate. a) q[1] is the control qubit, and q[0] is the target qubit. b) q[0] is the control qubit, and q[1] is the target qubit. The latter swaps amplitudes of |01⟩ and |11⟩ in the input vector [22]. Finally, CX$_{b)}$ = (H ⊗ H) CX$_{a)}$ (H ⊗ H), where "⊗" is the Kronecker product.

Incorporating the option selected above, if we multiply $F_{2^2}$ by itself, and we do the same with $F_{2^2}^{-1}$, then both multiplications result identical and also equal to the *CNOT* gate of Fig. 3(b) [1],

$$F_{2^2} \times F_{2^2} = F_{2^2}^{-1} \times F_{2^2}^{-1} = CNOT = \begin{bmatrix} 1 & 0 & 0 & 0 \\ 0 & 0 & 0 & 1 \\ 0 & 0 & 1 & 0 \\ 0 & 1 & 0 & 0 \end{bmatrix}, \qquad (8)$$

given that, multiplying $F_{2^2} \times F_{2^2}$ by $F_{2^2}^{-1} \times F_{2^2}^{-1}$ and regrouping, yields,

$$\left(F_{2^2} \times F_{2^2}\right) \times \left(F_{2^2}^{-1} \times F_{2^2}^{-1}\right) = F_{2^2} \times \left(F_{2^2} \times F_{2^2}^{-1}\right) \times F_{2^2}^{-1} = F_{2^2} \times I \times F_{2^2}^{-1} = F_{2^2} \times F_{2^2}^{-1} = I. \tag{9}$$

However,

$$\sqrt{CNOT} = \begin{bmatrix} 1 & 0 & 0 & 0 \\ 0 & (1+i)/2 & 0 & (1-i)/2 \\ 0 & 0 & 1 & 0 \\ 0 & (1-i)/2 & 0 & (1+i)/2 \end{bmatrix}, \tag{10}$$

therefore, considering Eq.(7), it is evident that $\sqrt{CNOT} \neq F_{2^2}$ and $\sqrt{CNOT} \neq F_{2^2}^{-1}$.

Based on Fig. 2, we can represent $F_{2^2} \times F_{2^2}$ and $F_{2^2}^{-1} \times F_{2^2}^{-1}$ in Fig. 4 through {Controlled-$S$, $H$} and {Controlled-$S^{-1}$, $H$} gates, respectively, which will help us to implement quantum entanglement and teleportation circuits in the next section.

a)
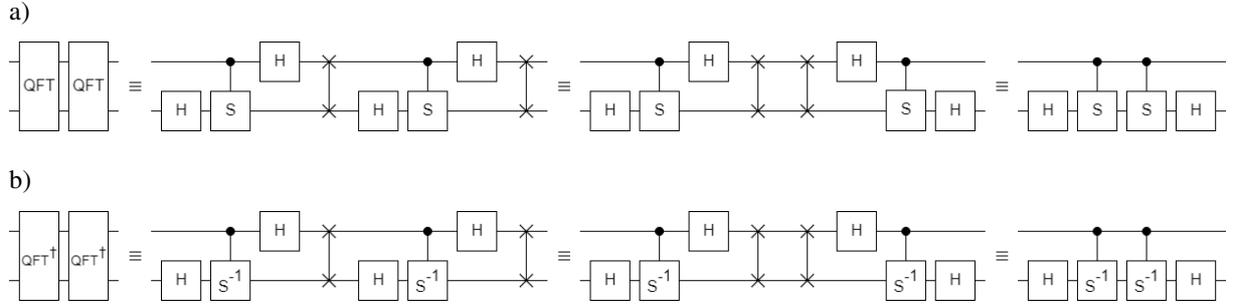
b)

FIG. 4. Representations of $F_{2^2} \times F_{2^2}$ and $F_{2^2}^{-1} \times F_{2^2}^{-1}$. a) $F_{2^2} \times F_{2^2}$ in terms of Controlled-$S$ and $H$ gates. b) $F_{2^2}^{-1} \times F_{2^2}^{-1}$ in terms of Controlled-$S^{-1}$ and $H$ gates.

Figure 5 shows the flipped *CNOT* gate, where the first equivalence is based on *H* and *CNOT* gates, while the second one is exclusively represented in terms of QFT and QFT$^{-1}$. Besides, Fig. 6 contains the *SWAP* gate equivalences in terms of {*H*, flipped *CNOT*, *CNOT*} gates, and {QFT, QFT$^{-1}$}.

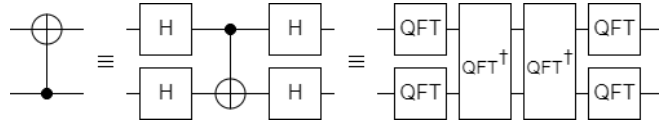

FIG. 5. Implementation of the flipped *CNOT* gate, first thanks to *H* and *CNOT* gates, and second one based on $F_{2^1} = H$ and $F_{2^2}^{-1} \times F_{2^2}^{-1} = CNOT$. In the literature, this gate is represented by the matrix of Fig. 3(b).

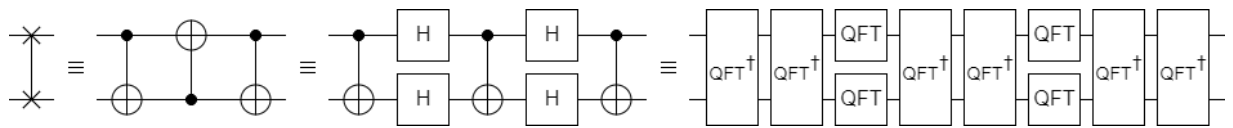

FIG. 6. Implementation of the *SWAP* gate, first one thanks to *CNOT* and flipped *CNOT* gates, second one thanks to *CNOT* and *H* gates, and finally based on $F_{2^1} = H$ and $F_{2^2}^{-1} \times F_{2^2}^{-1} = CNOT$.

If we now try to express each gate of the Controlled-*Gate* type by means of *CNOT* gates and the phase matrices of the previous section, and considering Eq.(8), we can start with the case of the Controlled-*S* and Controlled-*S*$^{-1}$ gates, where Fig. 7 shows their equivalences as a function of QFT$^{-1}$.

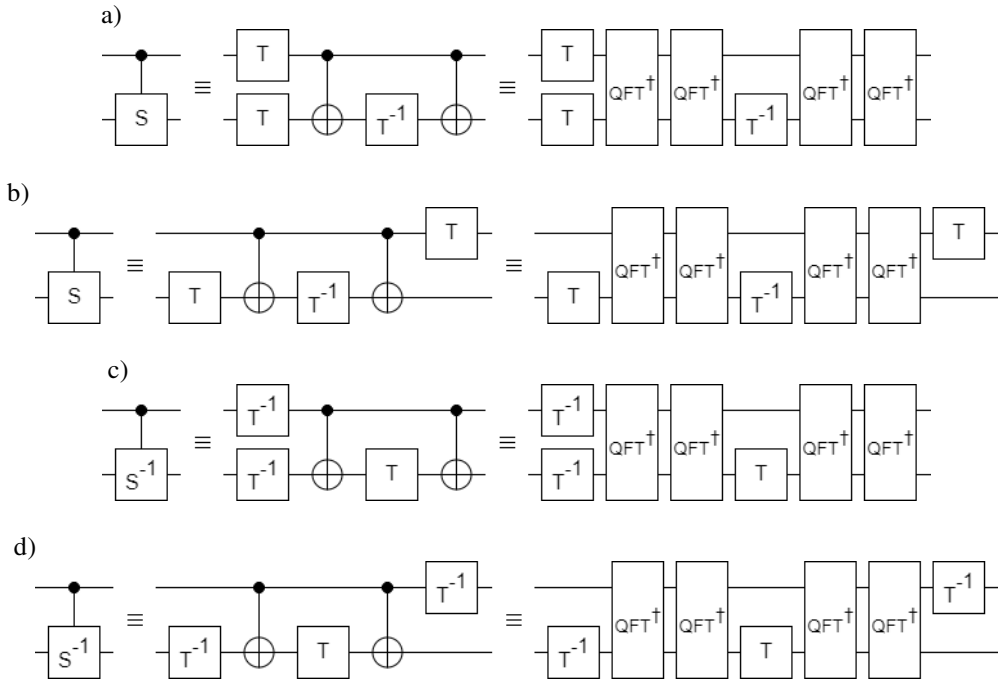

FIG. 7. Controlled-*S* and Controlled-*S*$^{-1}$ gates and their equivalences in terms of phase matrices, *CNOT* gate, and QFT$^{-1}$. a) First form of the Controlled-*S* in terms of $\{T, T^{-1}, CNOT\}$, and $\{T, T^{-1}, QFT^{-1} \times QFT^{-1}\}$. b) Second form of the Controlled-*S* in terms of $\{T, T^{-1}, CNOT\}$, and $\{T, T^{-1}, QFT^{-1} \times QFT^{-1}\}$. c) First form of the Controlled-*S*$^{-1}$ in terms of $\{T, T^{-1}, CNOT\}$, and $\{T, T^{-1}, QFT^{-1} \times QFT^{-1}\}$. d) Second form of the Controlled-*S*$^{-1}$ in terms of $\{T, T^{-1}, CNOT\}$, and $\{T, T^{-1}, QFT^{-1} \times QFT^{-1}\}$.

In Fig. 8, we implement the anti-control (or zero control) condition on a qubit being OFF for the *CNOT* gate [1-3], which constitutes a complement of the control condition on a qubit being ON that represents the original *CNOT* gate of Fig. 3.

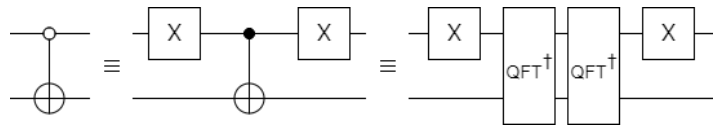

FIG. 8. Implementation of the anti-control (or zero control) condition on a qubit being OFF for the *CNOT* gate. The first equivalence thanks to *CNOT* gate and two inverters, while the last equivalence is in terms of QFT$^{-1}$.

Considering Fig. 4 and Eq.(8), we deduce that the *CNOT* gate has the equivalences present in Fig. 9.

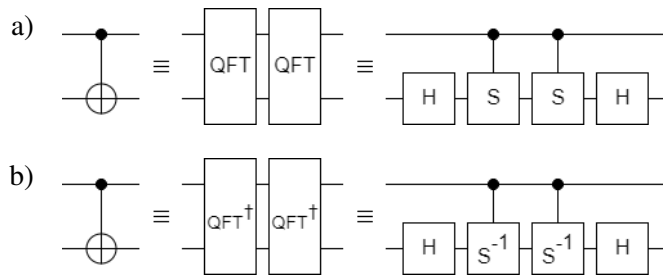

FIG. 9. Implementation of the *CNOT* gate. a) by QFT, or {*H*, Controlled-*S*}. b) by QFT$^{-1}$, or {*H*, Controlled-*S*$^{-1}$}.

From here on, and since each implementation that involves *CNOT* can be carried out both through QFT and its inverse, with the same criteria regarding the phase matrices {*S*, *T*, and so on}, then we will make the implementation of each case, exclusively based on QFT$^{-1}$. Then, we will implement the Controlled-*Z* gate, or *CZ*, which can be observed in Fig. 10 which its equivalences are in terms of {*H*, *CNOT*}, {QFT, QFT$^{-1}$}, {*H*, Controlled-*S*$^{-1}$}, and simply Controlled-*S*$^{-1}$. All these options will have an important role in the next section.

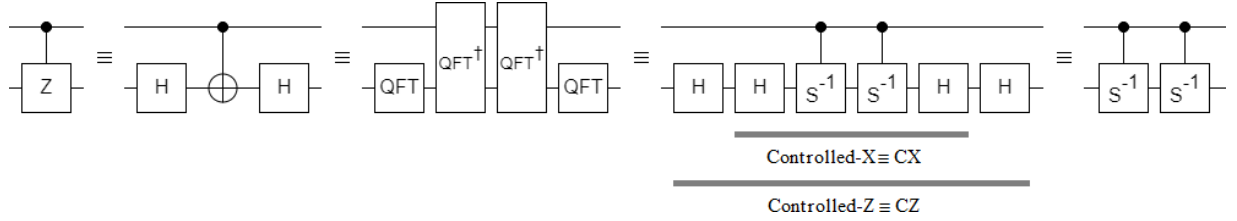

FIG. 10. Implementation of the Controlled-*Z* gate, where the first equivalence has to do with its representation via {*H*, *CNOT*}, the second one with the automatic replacement of the *H* and the *CNOT* gates with QFT and QFT$^{-1}$, respectively, while the second one used the version of the *CNOT* gate seen in Fig. 9(b). Finally, the last equivalence is a simplification of the previous case.

Figure 11 shows the Controlled-*Y* gate with its equivalences. The first one in terms of {*S*$^{-1}$, *CNOT*, *S*}, while the second one is obtained replacing *CNOT* with QFT$^{-1}$ × QFT$^{-1}$.

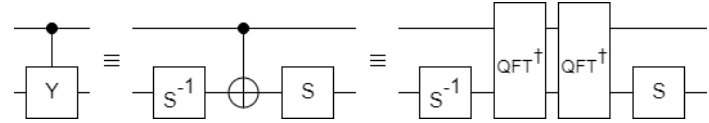

FIG. 11. Controlled-*Y* implemented thanks to {*S*$^{-1}$, *CNOT*, *S*} and {*S*$^{-1}$, $F_{2^2}^{-1} \times F_{2^2}^{-1}$, *S*}.

Figure 12 represents the Controlled-*T* gate, where the first equivalence is thanks to {*U*, *CNOT*, *U*$^{-1}$}, and the second one is via {*U*, $F_{2^2}^{-1} \times F_{2^2}^{-1}$, *U*$^{-1}$}.

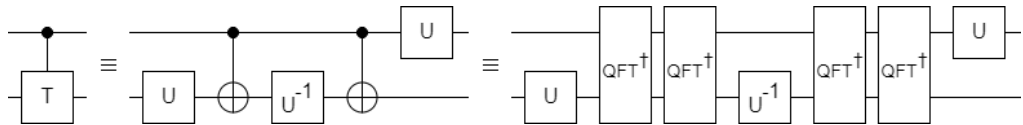

FIG. 12. Controlled-*T* implemented thanks to {*U*, *CNOT*, *U*$^{-1}$} and {*U*, $F_{2^2}^{-1} \times F_{2^2}^{-1}$, *U*$^{-1}$}.

The Controlled-*H* gate is implemented in Fig. 13 based on its equivalences and depends on Eq.(3). It could also be called Controlled-$F_2^1$, and can be represented via {*S*, *H*, *T*, *CNOT*, *T*$^{-1}$, *S*$^{-1}$} and {*S*, *H*, *T*, $F_{2^2}^{-1} \times F_{2^2}^{-1}$, *T*$^{-1}$, *S*$^{-1}$}, as we can see in Fig. 13.

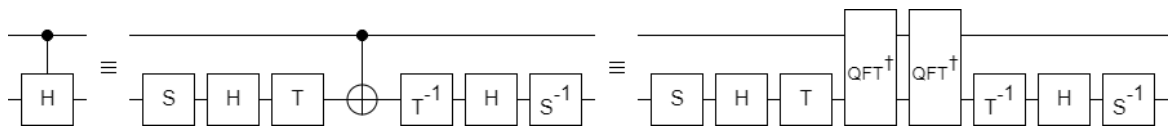

FIG. 13. Controlled-*H* implemented thanks to {*S*, *H*, *T*, *CNOT*, *T*$^{-1}$, *S*$^{-1}$} and {*S*, *H*, *T*, $F_{2^2}^{-1} \times F_{2^2}^{-1}$, *T*$^{-1}$, *S*$^{-1}$}.

Finally, Fig. 14 shows the implementation of the Controlled-*V* and Controlled-*V*$^{-1}$ gates in terms of {*H*, Controlled-*T*}, and {*H*, *T*, *T*$^{-1}$, flipped *CNOT*}; and {*H*, Controlled-*T*$^{-1}$}, and {*H*, *T*, *T*$^{-1}$, flipped *CNOT*}; respectively.

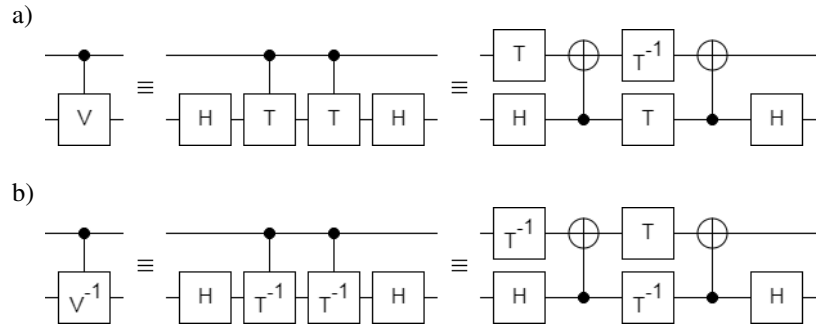

FIG. 14. Implementation of the Controlled-*V* and Controlled-*V*$^{-1}$ gates. a) by {*H*, Controlled-*T*}, or {*H*, *T*, *T*$^{-1}$, flipped *CNOT*}. b) by {*H*, Controlled-*T*$^{-1}$}, or {*H*, *T*, *T*$^{-1}$, flipped *CNOT*}.

In this section, we have implemented all versions of 2-qubit gates of the type Controlled-*Gate* expressed in terms of $CNOT = F_{2^2}^{-1} \times F_{2^2}^{-1}$, $H = F_{2^1}$, and the phase gates analyzed in the previous section, which, as we have seen, are derived from *H* gate.

### C. Equivalences for 3-qubit gates

With the same criteria used in Fig. 2 for the representation of QFT and QFT$^{-1}$ in the case of 2-qubits, Fig. 15 shows the equivalence between the QFT of 3-qubits and the gates {*H*, Controlled-*S*, Controlled-*T*, *SWAP*}, and between QFT$^{-1}$ and the gates {*H*, Controlled-*S*$^{-1}$, Controlled-*T*$^{-1}$, *SWAP*}.

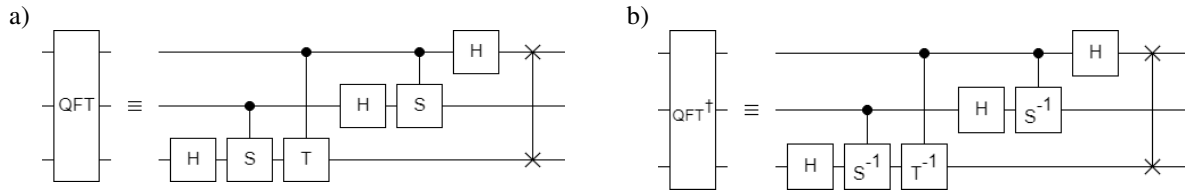

FIG. 15. a) QFT, $F_{2^3} \in \mathbb{C}^{2^3 \times 2^3}$ in terms of the {*H*, Controlled-*S*, Controlled-*T*, *SWAP*} gates. b) Inverse of QFT, $F_{2^3}^{-1} \in \mathbb{C}^{2^3 \times 2^3}$ in terms of the {*H*, Controlled-*S*$^{-1}$, Controlled-*T*$^{-1}$, *SWAP*} gates.

The implementation of $F_{2^3}^{-1} \times F_{2^3}^{-1}$, i.e., the 3-qubits QFT$^{-1}$ × QFT$^{-1}$ versions are represented in Fig. 16, where the first and second equivalences use {*H*, Controlled-*S*$^{-1}$, Controlled-*T*$^{-1}$, *SWAP*}, while the last one does not need the *SWAP* gate.

Figure 17 shows the techniques used to step over an intermediate qubit in a quantum circuit using the *CNOT* gate. All versions of Fig. 17 are equivalent, with more or less intervention of QFT$^{-1}$. In practice, the most recommendable of all the equivalences is undoubtedly the first one, however, all the versions together allow us to appreciate in detail the intervention of QFT$^{-1}$ behind a simple step over on a specific qubit. We must bear in mind, in this case, as in all the preceding ones, and in all those that follow, that the idea of presenting such an equivalence number does not have to do with a desperate search for alternatives, which is absurd since when we observe Fig. 17 in detail, it is evident that if we want to step over using QFT$^{-1}$×QFT$^{-1}$, with the first option is more than enough. The idea behind the presentation of all the alternatives has to do with the fact that whatever option is chosen, behind it will inevitably be Fourier. Moreover, as we have seen in Section II.A, all the phase matrices with which both QFT$^{-1}$ (Fig. 15) and QFT$^{-1}$×QFT$^{-1}$ (Fig. 16) are implemented are derived from Hadamard matrix *H*, which thanks to Eq.(3) is QFT = QFT$^{-1}$ for 1-qubit case.

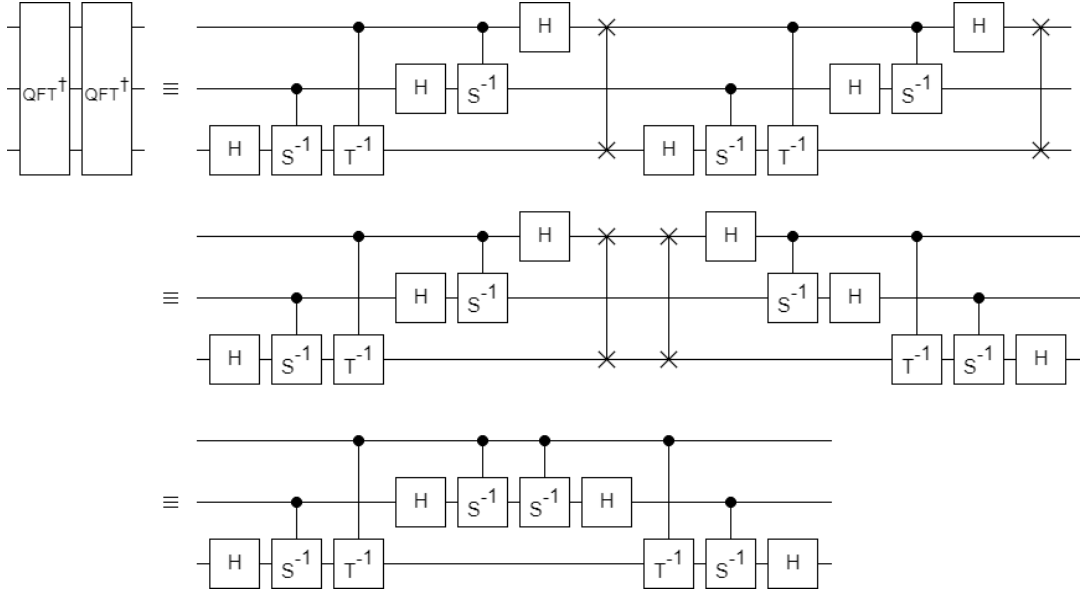

FIG. 16. Implementation of $F_{2^3}^{-1} \times F_{2^3}^{-1}$. The first and second equivalences use {$H$, Controlled-$S^{-1}$, Controlled-$T^{-1}$, SWAP}, while the last one only uses {$H$, Controlled-$S^{-1}$, Controlled-$T^{-1}$ }.

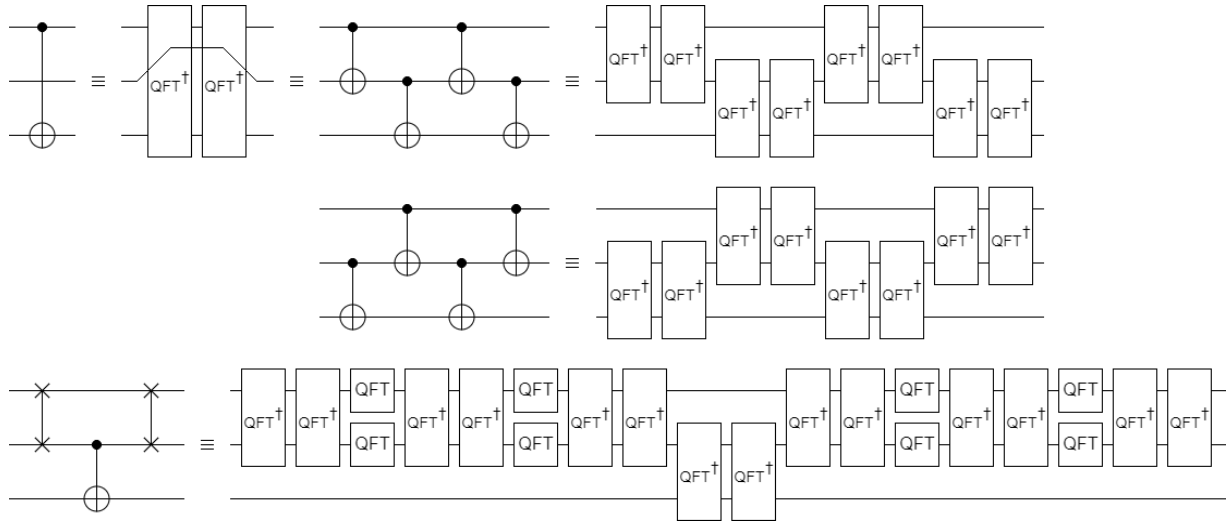

FIG. 17. Different equivalences for jumping an intermediate qubit, which range from the simplest and most practical version to the most complex. In all is present QFT$^{-1}$ × QFT$^{-1}$.

Figure 18 represents the double Feynman gate with its equivalences. This gate is fundamental for the generation of the Greenberger–Horne–Zeilinger state (GHZ state) [1-3], in this particular case, GHZ$_3$, which will be used to the maximum in the next section, where we will implement different configurations about entanglement [7-9] and teleportation [10-13]. The second line of Fig. 18 shows an implementation of the double Feynman gate in terms of one *CNOT* gate and one step over *CNOT* gate. Finally, in the same line, we have another implementation of it based on the equivalence of Fig. 9(b).

Figure 19 presents the first set of implementations of the 3-qubits gates: Toffoli, Fredkin, Peres, and Miller [23]. Fredkin, Peres, and Miller gates [23] are implemented based on the Toffoli, the step over *CNOT* (Fig. 17), and the flipped *CNOT* (Fig. 5) gates.

In Fig. 20 we have the second set of implementations of the 3-qubits gates: Toffoli, Fredkin, Peres, and Miller [23]. In this case, we replace the two known configurations of the Toffoli gate with the intervention of QFT$^{-1}$×QFT$^{-1}$. Instead, in the case of the Fredkin, Peres, and Miller gates we incorporate

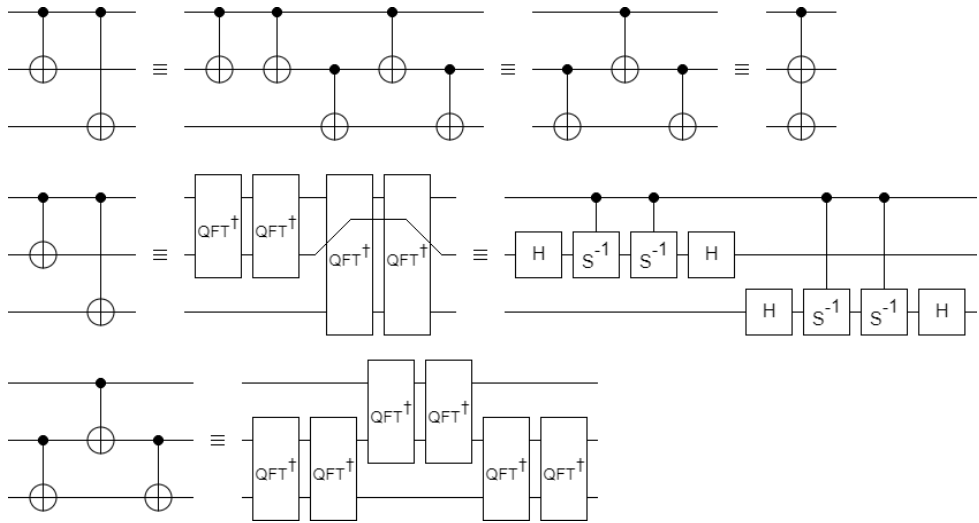

FIG. 18. Double Feynman gate and its equivalences. In the upper line, the step over configuration of Fig. 17 is used for the second *CNOT* gate, while every QFT$^{-1}\times$QFT$^{-1}$ is replaced by the equivalence of Fig. 9(b) in terms of the {$H$, Controlled-$S^{-1}$} gates.

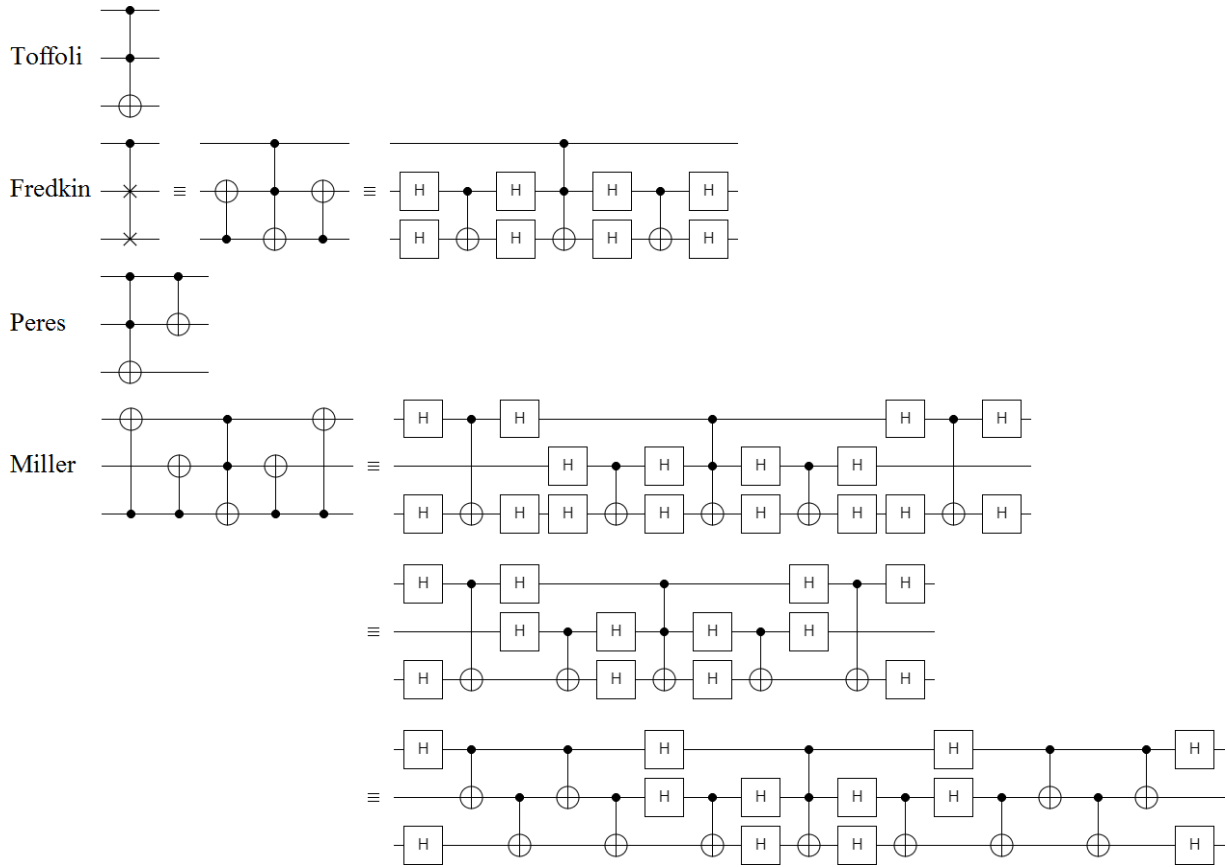

FIG. 19. Toffoli, Fredkin, Peres, and Miller gates, where the last three are implemented thanks to the Toffoli, the step over, and the flipped *CNOT* gates.

Toffoli gate in their interior as a known module. This greatly simplifies the implementation of these gates. Finally, all gates are fully expressed by QFT$^{-1}\times$ QFT$^{-1}$, i.e., Fourier.

Figure 21 shows the implementation of the Toffoli gate incorporating the version of QFT$^{-1}\times$ QFT$^{-1}$ in terms of the equivalence seen so far, i.e., based on {$H$, Controlled-$T$, Controlled-$S^{-1}$, Controlled-$T^{-1}$} gates, while Fredkin, Peres, and Miller in terms of {$H$, Controlled-$S^{-1}$, Toffoli as a module} gates.

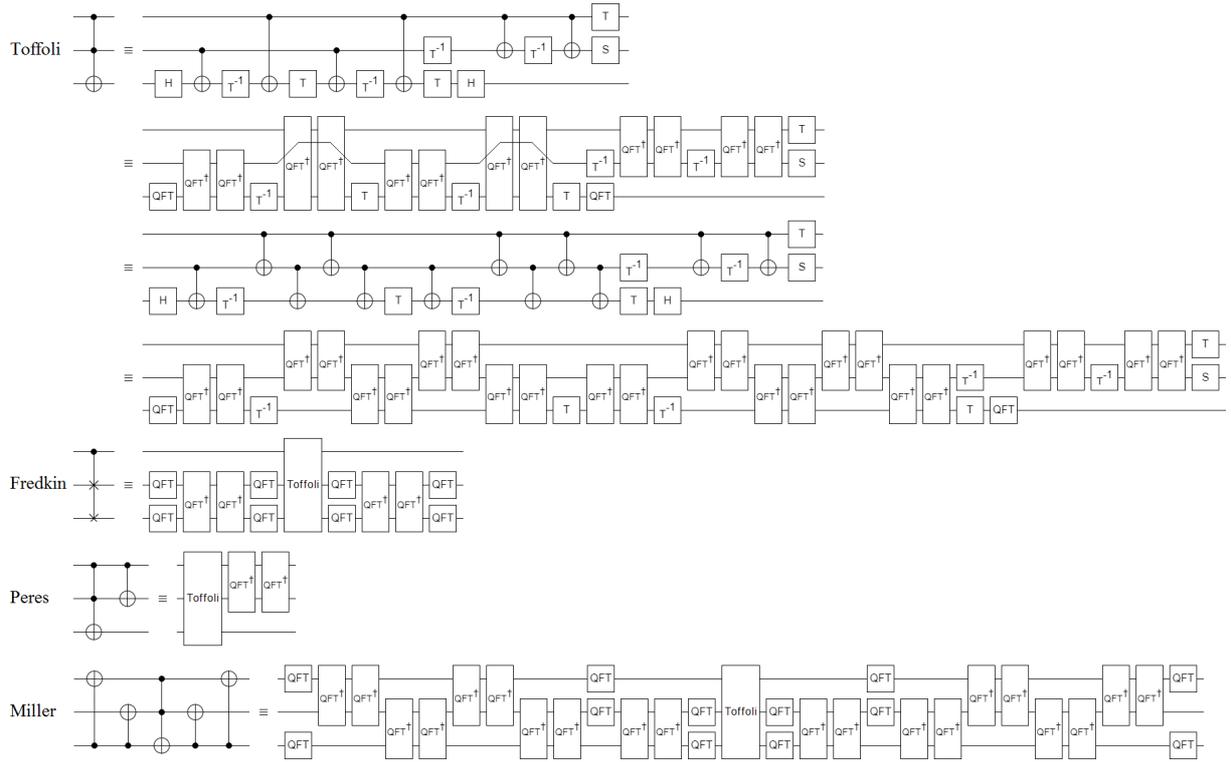

FIG. 20. Implementation of the Toffoli, Fredkin, Peres, and Miller gates thanks to QFT$^{-1}$ × QFT$^{-1}$. Additionally, the Fredkin, Peres, and Miller gates are implemented thanks to the Toffoli gate as a module.

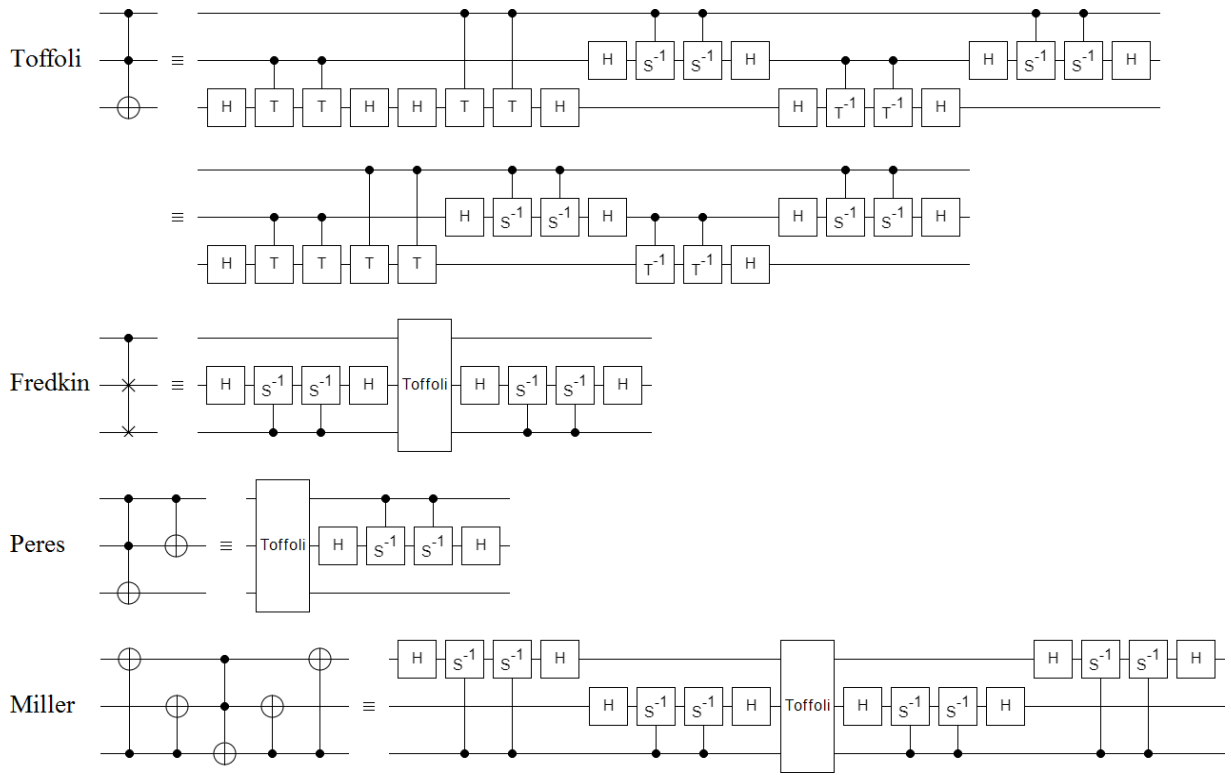

FIG. 21. Implementation of the Toffoli gate based on the {$H$, Controlled-$T$, Controlled-$S^{-1}$, Controlled-$T^{-1}$} gates, while the rest in terms of {$H$, Controlled-$S^{-1}$, Toffoli as a module} gates.

Figure 22 shows the Toffoli gate in terms of the gates $V = \sqrt{X}$ and $V^{-1}$ [1-3], their replacements according to Eq.(6), their simplifications, and, the replacement of the *CNOT* gate based on Fig. 9(b).

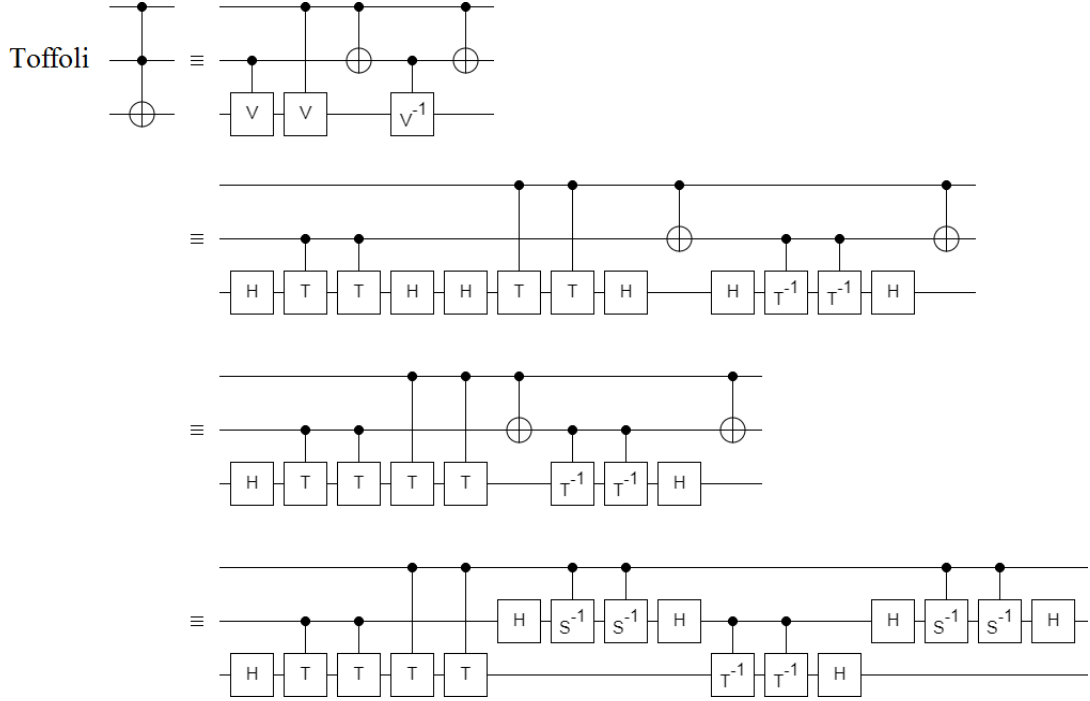

FIG. 22. Toffoli gate, where the first equivalence is implemented thanks to the $V$ and $V^{-1}$ gates. The second one in terms of the replacement of both gates according to Eq.(6). The second one corresponds to simplifications, while the last one has to do with the replacement of *CNOT* based on Fig. 9(b).

Finally, Fig. 23 represents the implementation of the Toffoli, Fredkin, Peres, and Miller gates thanks to a double Controlled-QFT$^{-1}$ gate, which notably simplifies the implementation of all the gates, since, here too the last three gates use the Toffoli gate like a module. We must take into account that each QFT$^{-1}$ matrix used in every Controlled-QFT$^{-1}$ gate is of the type $F_{2^2}^{-1}$, and no $F_{2^3}^{-1}$, then if $I_{2\times 2}$ is a 2×2 identity matrix, and $0_{2\times 2}$ is a 2×2 zero matrix, the Controlled-QFT-*1* gate will be,

$$Controlled - QFT^{-1} = \begin{bmatrix} I_{2\times 2} & 0_{2x2} \\ 0_{2x2} & F_{2^2}^{-1} \end{bmatrix}. \qquad (11)$$

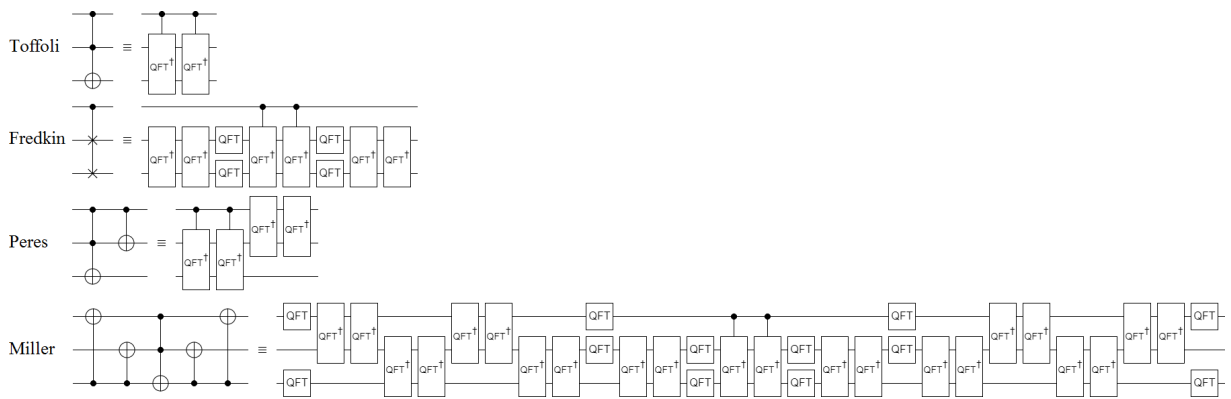

FIG. 23. The Toffoli, Fredkin, Peres, and Miller gates thanks to a simplification of the Toffoli gate based on a double Controlled-QFT$^{-1}$ gate.

## D. Equivalences for 4-qubit gates

As we have done in the previous cases, we present QFT and QFT$^{-1}$ for 4-qubits in Fig. 24, where the equivalences for QFT are based on the {$H$, Controlled-$S$, Controlled-$T$, Controlled-$U$, SWAP} gates, while, the equivalences for QFT$^{-1}$ depends on the {$H$, Controlled-$S^{-1}$, Controlled-$T^{-1}$, Controlled-$U^{-1}$, SWAP} gates.

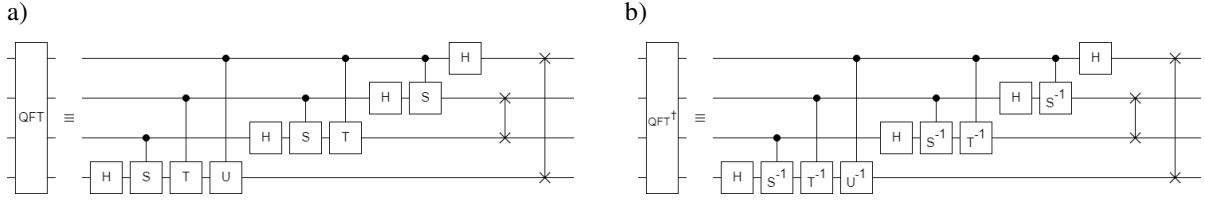

FIG. 24. a) QFT, $F_{2^4} \in \mathbb{C}^{2^4 \times 2^4}$ in terms of the {$H$, Controlled-$S$, Controlled-$T$, Controlled-$U$, SWAP} gates. b) Inverse of QFT, $F_{2^4}^{-1} \in \mathbb{C}^{2^4 \times 2^4}$ in terms of the {$H$, Controlled-$S^{-1}$, Controlled-$T^{-1}$, Controlled-$U^{-1}$, SWAP} gates.

In Fig. 25, we use the configuration of Fig. 24(b) to build QFT$^{-1}$×QFT$^{-1}$ for the 4-qubits case. These types of equivalences are very useful when we have to implement QFT, QFT$^{-1}$, and QFT$^{-1}$×QFT$^{-1}$ on platforms such as IBM Q Experience [16] or Rigetti [17] that do not have built-in versions of these tools in their respective toolboxes.

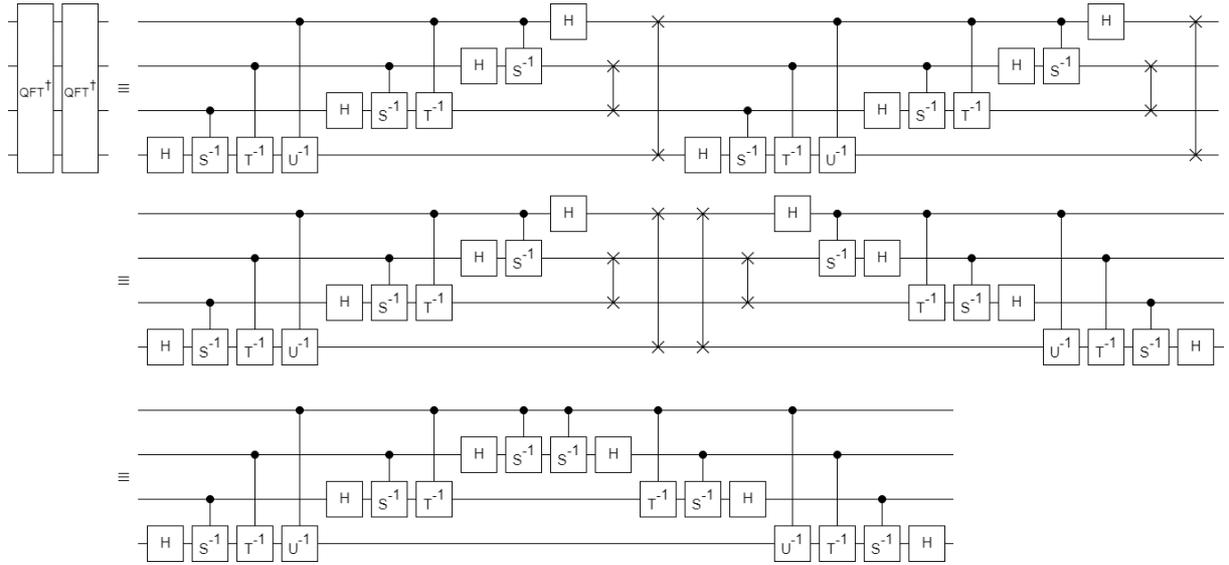

FIG. 25. QFT$^{-1}$×QFT$^{-1}$, first thanks to {$H$, Controlled-$S^{-1}$, Controlled-$T^{-1}$, Controlled-$U^{-1}$, SWAP}, and finally without the SWAP gate.

The triple Feynman gate is presented in Fig. 26. The first equivalence is based on one CNOT gate, and two step over CNOT gates, while the second one is obtained using the equivalence of Fig. 9(b).

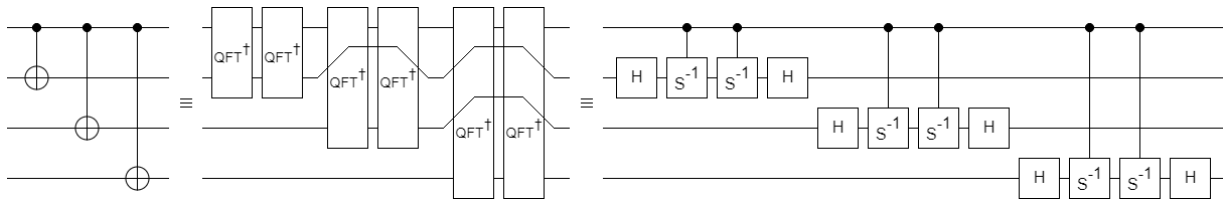

FIG. 26. Triple Feynman gate (with two step over CNOT gates) exclusively implemented thanks to QFT$^{-1}$×QFT$^{-1}$.

## III. QUANTUM ENTANGLEMENT, TELEPORTATION, AND SECRET SHARING

We will focus our analysis on the intervention of Fourier in Quantum Communications [24-26], by demonstrating its presence in Quantum Entanglement [7-9], Quantum Teleportation [10-13], and Quantum Secret Sharing [27]. Therefore, this section will make it clear that the two pillars of Quantum Information Processing (QIP): superposition and entanglement, are two attributes of unequivocally spectral characteristics.

### A. Quantum Entanglement

In this section, we will start with the so-called Bell states [1-3], which result from the combination of the Hadamard (*H*), and *CNOT* (Eq. 8) matrices, where,

$$|\beta_{00}\rangle = \frac{1}{\sqrt{2}}(|00\rangle + |11\rangle) = CX((I \otimes H)|00\rangle) = \begin{bmatrix} 1 & 0 & 0 & 0 \\ 0 & 0 & 0 & 1 \\ 0 & 0 & 1 & 0 \\ 0 & 1 & 0 & 0 \end{bmatrix} \left( \begin{bmatrix} 1 & 0 \\ 0 & 1 \end{bmatrix} \otimes \frac{1}{\sqrt{2}} \begin{bmatrix} 1 & 1 \\ 1 & -1 \end{bmatrix} \right) \begin{bmatrix} 1 \\ 0 \\ 0 \\ 0 \end{bmatrix} = \frac{1}{\sqrt{2}} \begin{bmatrix} 1 \\ 0 \\ 0 \\ 1 \end{bmatrix}. \quad (12)$$

The configuration of matrices of Eq.(12), which in optical circuits [15] is known as beamsplitter, is represented in Fig. 27 thanks to {*H*, *CNOT*}, {QFT, QFT$^{-1}$×QFT$^{-1}$}, and {*H*, Controlled-$S^{-1}$}. The latter allows us to obtain Bell states in terms of Fourier on platforms like IBM Q Experience [16], and Rigetti [17] which do not have QFT or QFT$^{-1}$ as built-in structures.

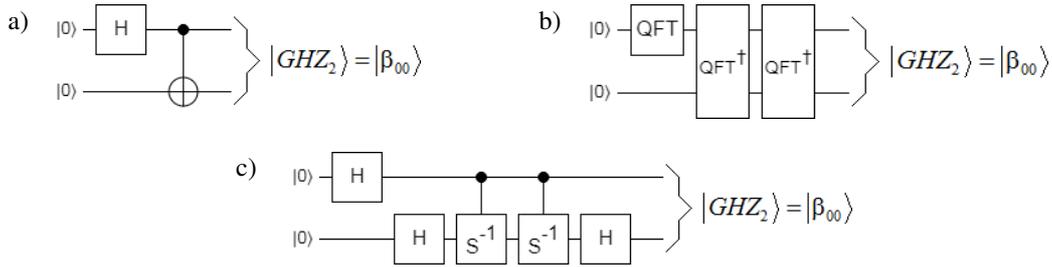

FIG. 27. Bell state via: a) {*H*, *CNOT*}, b) {QFT, QFT$^{-1}$×QFT$^{-1}$}, and c) {*H*, Controlled-$S^{-1}$}.

Finally, Fig. 28 shows GHZ$_3$ in terms of {*H*, *CNOT*, step over *CNOT*}, {QFT, QFT$^{-1}$×QFT$^{-1}$}, and {*H*, Controlled-$S^{-1}$}, while Fig. 29 does the same for GHZ$_4$. Both configurations (Figs. 28 and 29) will be of vital importance in the implementation of the Quantum Secret Sharing [27] protocol.

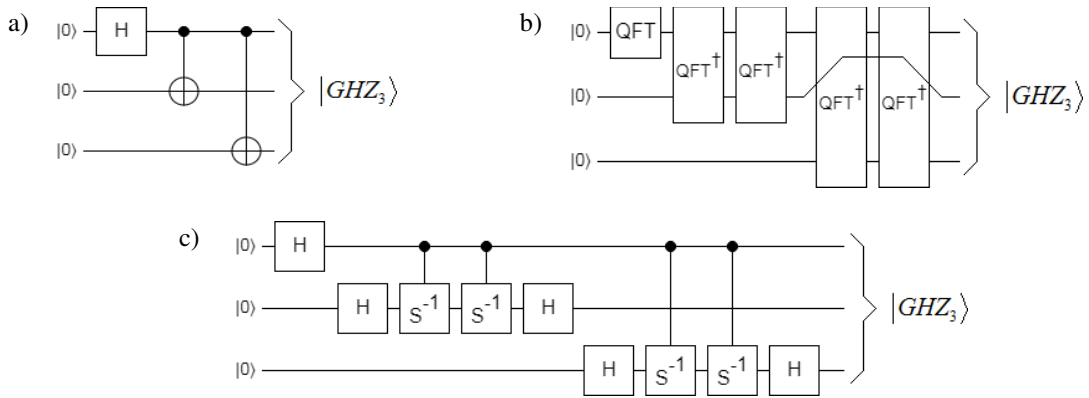

FIG. 28. GHZ$_3$ in terms of: a) {*H*, *CNOT*, step over *CNOT*}, b) {QFT, QFT$^{-1}$×QFT$^{-1}$}, and c) {*H*, Controlled-$S^{-1}$}.

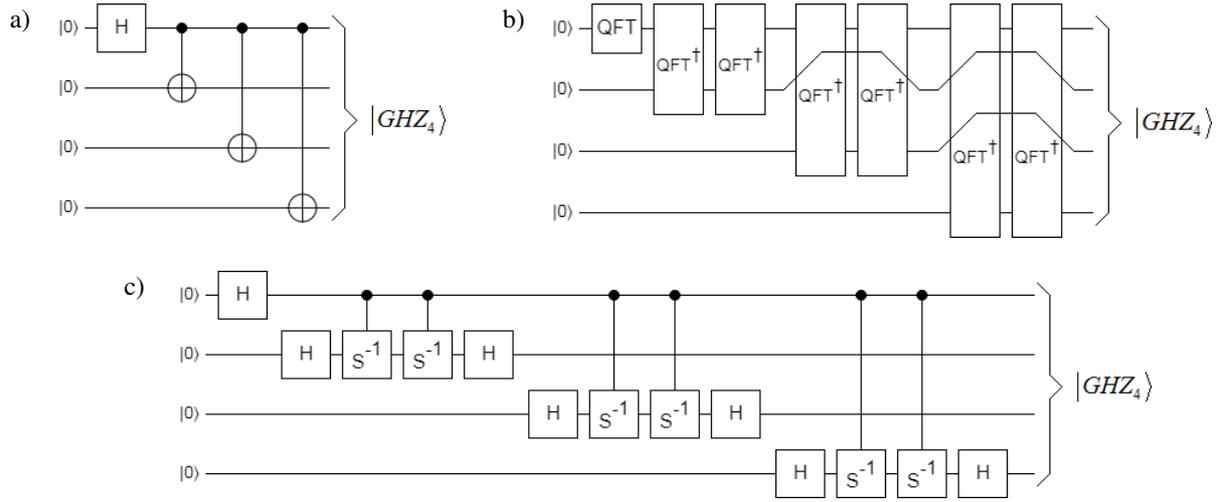

FIG. 29. GHZ$_4$ in terms of: a) {$H$, $CNOT$, step over $CNOT$}, b) {$QFT$, $QFT^{-1} \times QFT^{-1}$}, and c) {$H$, Controlled-$S^{-1}$}.

**B. Quantum Teleportation**

Using the equivalences of the immediately previous figures, we can implement the famous quantum teleportation protocol [10-13] in terms of Fourier. Figure 30 presents four options to carry out this task. The first is the one that we can find in all the literature that deals with Quantum Teleportation [10-13], which is based on {$H$, $CNOT$, Quantum Measurement (QuMe), $CZ$}. In the second one, we replace $CZ$ by {$H$, $CNOT$}. In the third one, we replace all the gates of the latter by their representation in terms of QFT, and QFT$^{-1}$, while in the last one, we implement QFT and QFT$^{-1} \times$QFT$^{-1}$ via the discrete gates that we can find in any commercial platform like IBM Q Experience [16], and Rigetti [17]. Finally, the pair {$H$, $CNOT$} is known as beamsplitter, and the group {$H$, $CNOT$, Quantum Measurement (QuMe)} is called Bell State Measurement (BSM), where the quantum measurement is carried out by Single-Photon Detectors (SPD) [11-13].

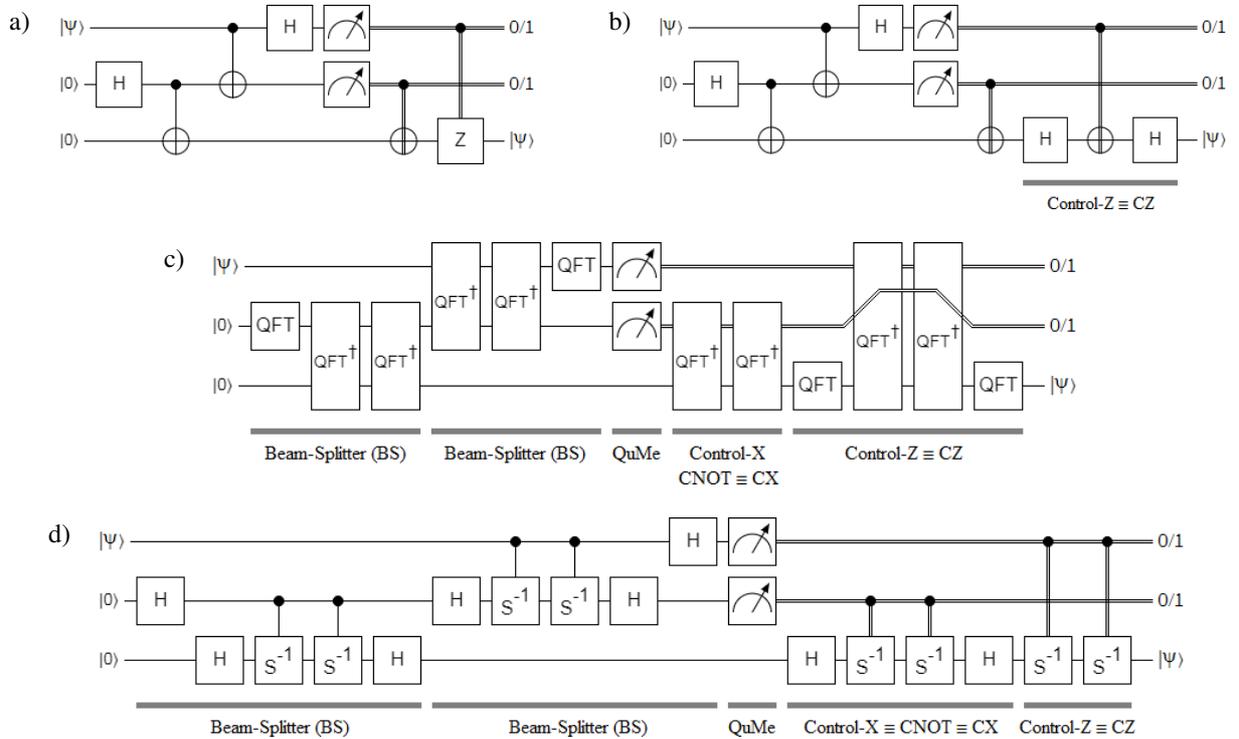

FIG. 30. Quantum Teleportation Protocol: a) the traditional one, b) replacing CZ by its equivalence based on {$H$, $CNOT$}, c) in terms of {QFT, QFT$^{-1} \times$QFT$^{-1}$, QuMe}, and d) via {$H$, Controlled-$S^{-1}$, QuMe}.

## C. Quantum Secret Sharing

Among other things, the quantum secret sharing (QSS) protocol [27] is a generalization of the quantum teleportation protocol previously seen for the case where we worked with GHZ [1-3] instead of Bell states. Figure 31 shows five different implementations of QSS for $GHZ_3$ thanks to $\{H, CNOT,$ step over $CNOT$, QuMe, $CZ\}$, $\{H, CNOT,$ step over $CNOT$, QuMe$\}$, $\{QFT, QuMe, QFT^{-1} \times QFT^{-1}\}$, and $\{H,$ Controlled-$S^{-1}$, QuMe$\}$. With the same criteria, Fig. 32 shows the same for $GHZ_4$.

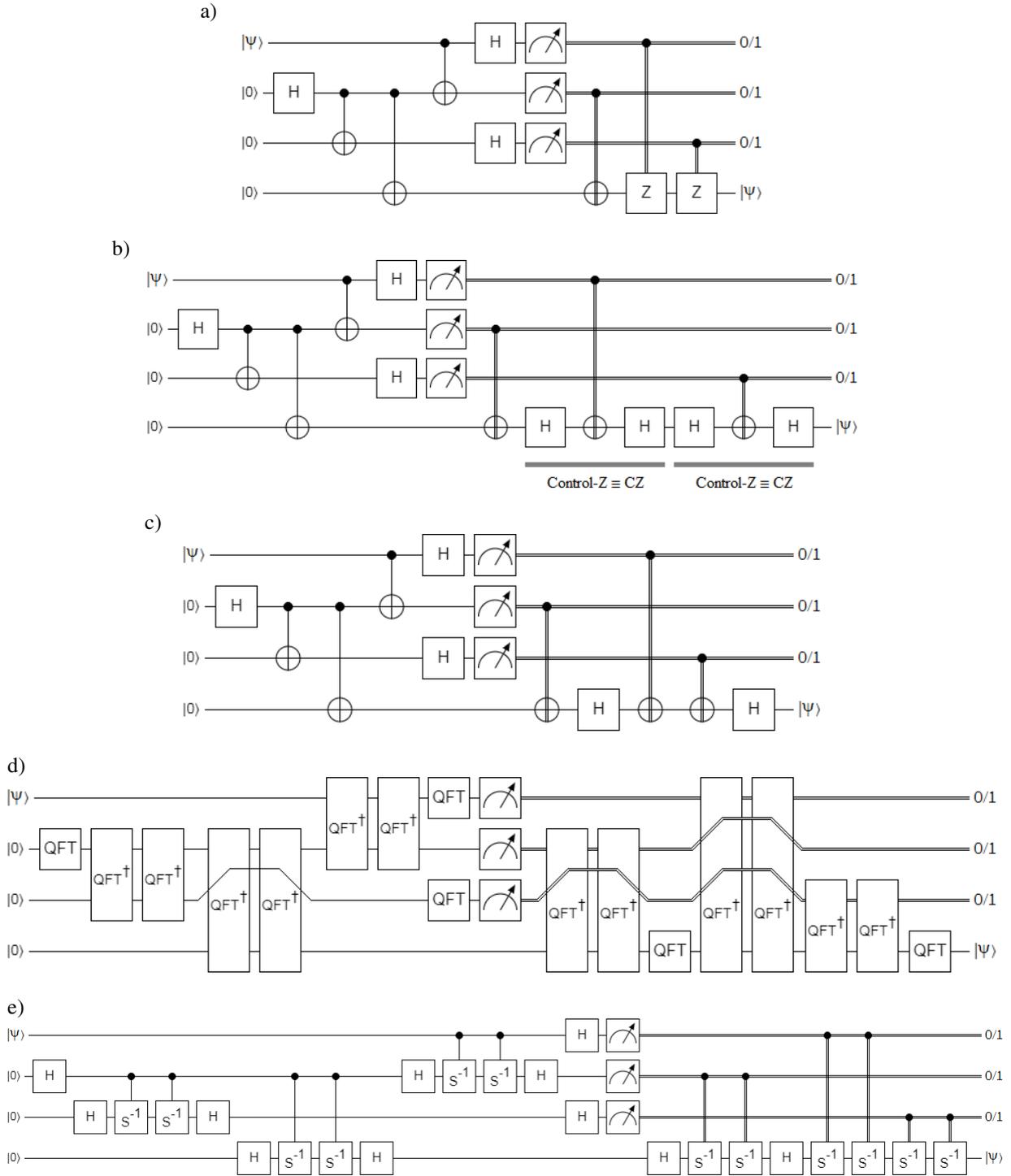

FIG. 31. Quantum Secret Sharing protocol for $GHZ_3$. a) The traditional one. b) $CZ$ is replaced by $\{H, CNOT\}$. c) Simplification of the last case. d) Replacement of each gate of the previous case with $\{QFT, QFT^{-1} \times QFT^{-1}\}$. e) Implementation of $\{QFT, QFT^{-1} \times QFT^{-1}\}$ thanks to $\{H,$ Controlled-$S^{-1}\}$.

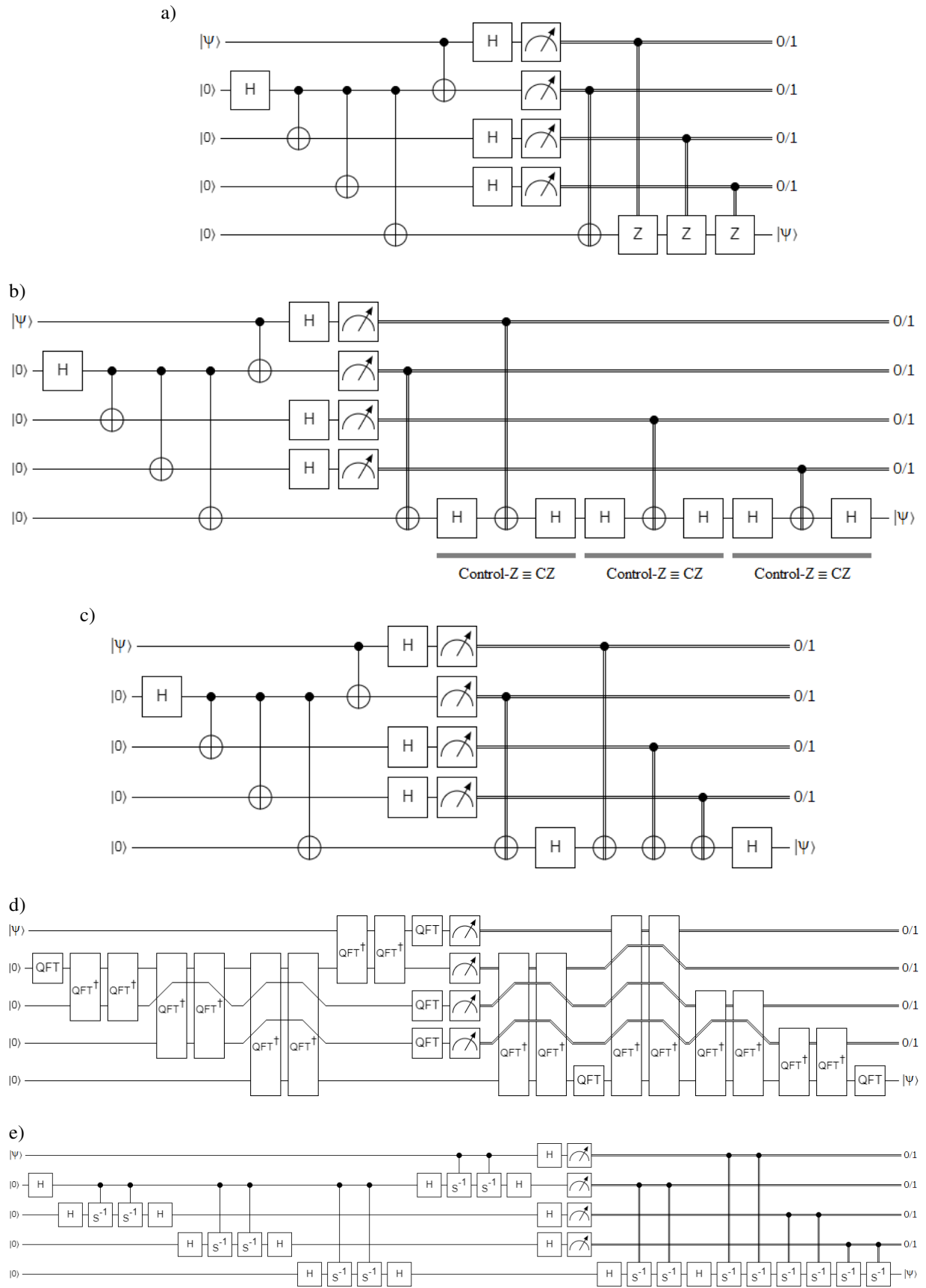

FIG. 32. Quantum Secret Sharing protocol for GHZ$_4$. a) The traditional one. b) *CZ* is replaced by {*H, CNOT*}. c) Simplification of the last case. d) Replacement of each gate of the previous case with {QFT, QFT$^{-1}\times$QFT$^{-1}$}. e) Implementation of {QFT, QFT$^{-1}\times$QFT$^{-1}$} thanks to {*H*, Controlled-$S^{-1}$}.

# IV. CONCLUSIONS

We have demonstrated that quantum information processing (QIP) completely rests on quantum Fourier transform (QFT). In fact, every gate, circuit, algorithm, procedure, and QIP protocol can be expressed thanks to QFT and its inverse. On the other hand, we know that the Pauli matrices are the fundamental building blocks of QIP [1-3], given that they form a complete basis for any Hermitian operator. In consequence, any observable can be expanded into strings of Pauli operators, the expectation values of which we can measure efficiently with a quantum computer. Section II.A highlighted the close link between the Pauli and Hadamard matrices. Therefore, this demonstrates the unequivocal presence of Fourier, 190 years after his death, as a substrate for this novel discipline.

Finally, this study makes clear the spectral nature of quantum entanglement [7-9], quantum teleportation [10-13], and therefore quantum secret sharing [27], with an obvious projection on quantum communications [24-26], in general, and quantum Cryptography [28, 29] as well as quantum Internet [30-34], in particular.

# ACKNOWLEDGEMENTS

M.M. thanks to Prof. S.S. Iyengar, director of the School of Computing and Information Sciences of Florida International University for all his help and support, Prof. Frank Rioux, Professor Emeritus, Department of Chemistry, St. John's University and College of St. Benedict, for his excellent comments on the intervention of the Pauli's matrices as the fundamental building blocks of Quantum Information Processing.

---

1. M. A. Nielsen and I. L. Chuang, *Quantum Computation and Quantum Information*, 3rd ed. (Cambridge University Press, Cambridge, U.K., 2004).
2. P. Kaye, R. Laflamme and M. Mosca, *An Introduction to Quantum Computing* (Oxford University Press, Oxford, U.K., 2004).
3. J. Stolze and D. Suter, *Quantum Computing: A Short Course from Theory to Experiment* (Wiley, Weinheim, Germany, 2007).
4. J. Chiaverini, *et al*., Science 308, 997 (2005).
5. Y. S. Weinstein, *et al*., Phys. Rev. Lett. 86, 1889 (2001).
6. Y. S. Weinstein, *et al*., J. Chem. Phys. 121, 6117 (2004).
7. J. Audretsch, *Entangled Systems: New Directions in Quantum Physics* (Wiley, Weinheim, Germany, 2007).
8. G. Jaeger, *Entanglement, Information, and the Interpretation of Quantum Mechanics*. The Frontiers Collection (Springer-Verlag, Berlin, Germany, 2009).
9. R. Horodecki, *et al*., Rev. Mod. Phys. 81, 865 (2009).
10. C. H. Bennett, *et al*., Phys. Rev. Lett. 70, 1895 (1993).
11. D. Bouwmeester, *et al*., Nature 390, 575 (1997).
12. B. D. Bouwmeester, *et al*., Phil. Trans. R. Soc. Lond. A 356, 1733 (1998).
13. D. Boschi, *et al*., Phys. Rev. Lett. 80, 1121 (1998).
14. P. W. Shor, SIAM J. Comp. 26, 1484 (1997).
15. A. Furusawa and P. van Loock, *Quantum Teleportation and Entanglement: A Hybrid Approach to Optical Quantum Information Processing* (Wiley, Weinheim, Germany, 2011).
16. IBM Q Experience [https://quantum-computing.ibm.com/]
17. Rigetti Computing [https://www.rigetti.com/]
18. D. E. Browne, New J. Phys. 9, 146 (2007).
19. E. Rieffel and W. Polak, *Quantum Computing: A Gentle Introduction* (The MIT Press, Cambridge, Massachusetts, 2011).
20. A. Fijany and C. P. Williams, arXiv:9809004.
21. G. Kurizki and G. Gordon, *The Quantum Matrix: Henry Bar's Perilous Struggle for Quantum Coherence*, pp.136-138 (Oxford University Press, Oxford, UK, 2020).
22. Qiskit: Summary of Quantum Operations

[https://qiskit.org/documentation/tutorials/circuits/3_summary_of_quantum_operations.html]
23. L. Perkowski, Introduction and Quantum Mechanics [http://web.cecs.pdx.edu/~mperkows/CLASS_FUTURE/NEW_MATERIALS_2011/lukac_perkowski_book_introduction_and_quantum_mechanics.pdf]
24. G. Cariolaro, *Quantum Communications: Signals and Communication Technology* (Springer, Switzerland, 2015).
25. S. Imre and L. Gyongyosi, *Advanced Quantum Communications: An Engineering Approach* (Wiley-IEEE Press, N.Y., USA, 2013).
26. M. Benslama, A. Benslama and S. Aris, *Quantum Communications in New Telecommunications Systems* (John Wiley & Sons, Inc., Hoboken, USA, 2017).
27. D. Joy, M. Sabir, B.K. Behera and P.K. Panigrahi, Implementation of quantum secret sharing and quantum binary voting protocol in the IBM quantum computer. Quant. Info. Proc., 19:33 (2020).
28. .-, *Lecture Notes in Physics 797: Applied Quantum Cryptography*. C. Kollmitzer and M. Pivk Eds. (Springer, Heidelberg, Germany, 2010).
29. Ekert, A.: Quantum Cryptography, in A.V. Sergienko (Ed.): Quantum Communications and Cryptography: Optical Science and Engineering (CRC Press, Boca Raton, USA, 2006).
30. W. Dür, R. Lamprecht and S. Heusler, Towards a quantum internet, Eur. J. Phys., 38, 043001 (2017).
31. H. J. Kimble, The quantum internet, Nature, 453, 1023–1030 (2008).
32. L. Gyongyosi and S. Imre, Entanglement Accessibility Measures for the Quantum Internet, Quant. Info. Proc., 19:115 (2020).
33. L. Gyongyosi and S. Imre, Entanglement Access Control for the Quantum Internet. arXiv: 1905.00256.
34. L. Gyongyosi and S. Imre, Opportunistic Entanglement Distribution for the Quantum Internet. arXiv:1905.00258.